\documentclass [6pt,a4paper]{article}
\usepackage{bbm}
\usepackage{amsfonts}
\usepackage{mathrsfs}
\usepackage {amssymb}
\usepackage {amsmath}
\usepackage{amsthm}
\usepackage{latexsym}
\usepackage{booktabs}
\def\dse#1{\vskip 0.6cm\noindent
        {\large\bf #1}
        \vskip 0.4cm}

\usepackage{lastpage}
\usepackage{epsfig}

\begin{document}
\begin{center}
\textbf{\large{Some results of linear codes over the ring
$\mathbb{Z}_4+u\mathbb{Z}_4+v\mathbb{Z}_4+uv\mathbb{Z}_4$ }}\footnote { E-mail
addresses:
 lpmath@126.com(P.Li), gxmhgd@126.com(X.Guo).\\
}
\end{center}

\begin{center}
{ { Ping Li,  Xuemei Guo, Shixin Zhu} }
\end{center}

\begin{center}
\textit{School of Mathematics, Hefei University of
Technology, Hefei 230009, Anhui, P.R.China \\
 }
\end{center}

\noindent\textbf{Abstract:} In this paper, we mainly study the theory of linear codes over the ring
$R =\mathbb{Z}_4+u\mathbb{Z}_4+v\mathbb{Z}_4+uv\mathbb{Z}_4$. By the Chinese Remainder Theorem, we have $R$ is isomorphic to the direct sum of four rings $\mathbb{Z}_4$. We define a Gray map $\Phi$ from $R^{n}$ to $\mathbb{Z}_4^{4n}$, which is a distance preserving map. The Gray image of a cyclic code over $R^{n}$ is a linear code over $\mathbb{Z}_4$. Furthermore, we study the MacWilliams identities of linear codes over $R$ and give the the generator polynomials of cyclic codes over $R$. Finally, we discuss some properties of MDS codes over $R$.

\noindent\emph{Keywords}: Linear code; Gray map; MacWilliams identities; Cyclic codes; MDS codes.
\dse{1~~Introduction}

Since Hammons et al.[1] have showed certain good binary nonlinear codes (Kerdock codes and Preparata codes) are actually images of some linear codes over $\mathbb{Z}_4$ via the Gray map, which has made a breakthrough in coding theory. In recent twenty years, linear codes over finite chain rings and some special finite non-chain rings have been extensively studied especially about the structure of cyclic codes and constacyclic codes. For example, Dinh et al.[2] have considered the structure of cyclic and negacyclic codes over finite chain rings. Blackford[3] has studied the structure of negacyclic codes of arbitrary lengths over $\mathbb{Z}_4$. Li et al.[4] have studied the structure of cyclic codes of arbitrary lengths over $\mathbb{F}_q  + u\mathbb{F}_q$. Zhu et al.[5] have began to construct good codes over the quaternary commutative non-chain rings and also studied the structure and properties of a class of constacyclic codes over $\mathbb{F}_p  + v\mathbb{F}_p$($v^2=v$). Since then, the encoding issues about non-chain expansion of finite chain rings interested many researchers. Yildiz et al. [7,8] have expanded the ring $\mathbb{F}_2  + u\mathbb{F}_2$ ($u^2 = 0$) to $\mathbb{F}_2+u\mathbb{F}_2+v\mathbb{F}_2+uv\mathbb{F}_2$($u^{2}=0$,$v^{2}=0$,$uv=vu$) and constructed some optimal codes. At the same time, Kai et al.[9] have got good binary codes via Gray map. Yildiz et al.[10] have constructed the formally self-dual codes over $\mathbb{Z}_4+u\mathbb{Z}_4$ and got the formally self-dual codes over $\mathbb{Z}_4$. Gao et al. [11] have studied the linear codes over $\mathbb{Z}_4+v\mathbb{Z}_4$($v^{2}=v$). Bandi et al.[12] have constructed the self-dual codes over $\mathbb{Z}_4+v\mathbb{Z}_4$($v^{2}=v$) and given the relation between self-dual codes over $\mathbb{Z}_4+v\mathbb{Z}_4$($v^{2}=v$) and $\mathbb{Z}_4$. In [13], they have studied the structure of a class of constacyclic codes of arbitrary lengths.

MacWilliams identity is an useful tool in studying weight distributions of linear codes and their duals. Recently, the study of the MacWilliams identities of linear codes over finite rings has been a hot topic in encoding theory. Zhu. et al.[14] have studied the symmetrical MacWilliams identities over $\mathbb{Z}_{k}$. The weight distribution and MacWilliams identities of linear codes over finite chain and non-chain rings interested more encoding researchers. Li. et al.[15] have studied a type of MacWilliams identity for linear codes over $\mathbb{Z}_4+u\mathbb{Z}_4$ on Lee weight.

Maximum-distance separable(MDS) codes over finite fields are very important in coding theory and have been studied extensively. Recently, a number of papers have been published dealing with related codes over finite rings. In[15], Dougherty and Shiromoto studied maximum distance with respect to rank codes over $\mathbb{Z}_{k}$,i.e., linear codes $C$ of length $n$ with minimum Hamming weight equal to $n-rank(C)+1$. Shiromoto proved a bound on the minimum general weights for codes over finite commutative rings with to the respect to the orders of codes. In this work, we examine Singleton bound on Hamming weights for codes over $R$.

In this paper, we consider another non-chain expansion of $\mathbb{Z}_4$, which is the ring $\mathbb{Z}_4+u\mathbb{Z}_4+v\mathbb{Z}_4+uv\mathbb{Z}_4$($u^{2}=u$,$v^{2}=v$,$uv=vu$). Here we study linear codes and  their corresponding properties over this ring.

\dse{2~~Preliminaries}
Let $R=\mathbb{Z}_4+u\mathbb{Z}_4+v\mathbb{Z}_4+uv\mathbb{Z}_4$, where $u^{2}=u$,$v^{2}=v$,$uv=vu$. Note that the ring $R$ can also be viewed as the quotient ring $\frac{\mathbb{Z}_4[u,v]}{(u^2,v^2,uv-vu)}$. Let $r$ be any element of $R$, which can be expressed uniquely as $r=a+ub+cv+uvd$, where $a$,$b$,$c$,$d\in\mathbb{Z}_4$. Let $(R,+)$ be an $R-$additive group, and $(R, + )\cong(\mathbb{Z}_4,+)\otimes(\mathbb{Z}_4,+)\otimes(\mathbb{Z}_4,+)\otimes(\mathbb{Z}_4,+)$, where $(\mathbb{Z}_4,+)$ is $\mathbb{Z}_4-$additive group. Let $(R,\ast)$ be a $R-$unit group, and $(R,\ast)\cong(\mathbb{Z}_4,\ast)\otimes(\mathbb{Z}_4,\ast)\otimes(\mathbb{Z}_4,\ast)\otimes (\mathbb{Z}_4,\ast)$, where $(\mathbb{Z}_4,\ast)$ is $\mathbb{Z}_4-$unit group.

An element $e$ is called an idempotent element if $e^{2}=e$ .

Let $x$,$y$ be any two elements over $R$,  then we define the Euclidean inner-product on $R$  by taking $x\cdot y$. If $x\cdot y =0$, then $x$ and $y$ are called orthogonal.

Let $e_1=1-u-v+uv$, $e_2=u-uv$, $e_3=v-uv$, $e_4=uv$, then $e_1,e_2,e_3,e_4$ are pairwise orthogonal non-zero idempotent elements over $R$, and the unit element 1 can be decomposed as $1=e_1+e_2+e_3+e_4$. By the Chinese Remainder Theorem, we have $ R=e_1R+e_2R+e_3R+e_4R$, and $r$ can be expressed uniquely as $r = r_1e_1+r_2e_2+r_3e_3+r_4e_4$, where $r_1=a$,$r_2=a+b$,$r_3=a+c$,$r_4=a+b+c+d$. Here we define a $\mathbb{Z}_4-$linear map $\phi:r\mapsto(r_1,r_2,r_3,r_4)$. We expand $\phi$ as:
\begin{align*}
\Phi :R^n &\rightarrow \mathbb{Z}^{4n}_4 \\
(c_0,c_1,\cdots,c_{n-1})& \mapsto(r_{1,0},\cdots,r_{1,n-1},r_{2,0},\cdots,r_{2,n-1},r_{3,0},\cdots,r_{3,n-1},r_{4,0},\cdots,r_{4,n-1})
\end{align*}
where $c_i\in R$, and $\Phi$ is called a Gray map over the ring $R$.\\

Let $C$ be a $R$-submodule over $R^n$, then $C$ is called a linear code of length $n$ over $R$. We define a cyclic shift operator as: \[\tau(c_0,c_1,\cdots,c_{n-1})=(c_{n-1},c_0,\cdots,c_{n-2}).\]

If any $\textbf{c}\in C$, we have $\tau(\textbf{c})\in C$. Then $C$ is called a cyclic code over $R$. Let $\textbf{c}=(c_0,\cdots,c_{n - 1})\in C$, which is equivalent to $c(x)=\sum_{i=0}^{n-1}{c_ix^i}$ under an isomorphic map, then $C$ is a cyclic code if and only if $C$ is an ideal of $R[x]/(x^n-1)$. Define \[C^\bot=\{\textbf{x} \in R^n |\textbf{x}\cdot \textbf{y}= 0,\forall ~\textbf{y}\in C\},\]
which is called the dual code of $C$. Clearly, $C^{\bot}$ is also a linear code over $R$. A code $C$ is said to be self- orthogonal if $C\subseteq C^\bot$, and self-dual if $C=C^\bot$.

The Lee weight of $0,1,2,3\in\mathbb{Z}_4$,  denoted by $w_L(0),w_L(1),w_L(2),w_L(3)$
respectively, are defined by $w_L(0)=0$,$w_L(1)=w_L (3)=1$,$w_L({\rm{2}})={\rm{2}}$. The Lee weight $w_L(r^{\prime})$ of $r^{\prime}=(r_1,r_2,r_3,r_4)\in\mathbb{Z}_4^4$ is defined to be the integral sum of the Lee weight of its components: $$w_L (r^{\prime})=w_L(r_1,r_2,r_3,r_4)=\sum_{i = 1}^4{w_L (r_i )}.$$ For any $\textbf{a}=(a_0,\cdots,a_{n-1} ),~\textbf{b}=(b_0,\cdots,b_{n-1})\in R^n$, the Lee weight of $\textbf{a}$ is denoted by $w_L(\textbf{a})=\sum_{i=0}^{n-1}{w_L(a_i)}$, and the Lee distance of $\textbf{a},~\textbf{b}$ is denoted by $d_L (\textbf{a},\textbf{b})=w_L(\textbf{a}-\textbf{b})$. The Lee distance of $C$ is defined by \[ d_L(C)=\min \{d_L(\textbf{a}-\textbf{b}),\forall~\textbf{a},\textbf{b}\in C,~\textbf{a} \neq \textbf{b}\}.\]
If $C$ is a linear code, then $d(C)$ is the minimum Lee weight of non-zero codewords in $C$. And $C$ can be expressed as $(N,M,d)$, where $M, N, d$ are the length, the number of the codewords and the minimum Lee distance of $C$, respectively.

We define the Hamming weight as the number of non-zero coordinates of $\textbf{c}$, where
$\textbf{c}=(c_{0},c_{1},\cdots,c_{n-1})$. The Hamming distance of $C$ is defined by
\[ d_H(C)=\min \{d_H(\textbf{a}-\textbf{b}),\forall~ \textbf{a},\textbf{b}\in C,~\textbf{a} \neq \textbf{b}\}.\]
For $C$ is a linear code, then $d_{H}(C)$ is the minimum Hamming weight of non-zero codewords in $C$.\\

\dse{3~~Linear codes over $R$}

 For any $\textbf{r}=(r^{(0)},r^{(1)},\cdots,r^{(n-1)})\in R^n$, where $\textbf{r}^{(i)}=r_{i1}e_1+r_{i2}e_2+r_{i3}e_3+r_{i4}e_4$ and $i=0,1,\cdots,n-1$. Then $\textbf{r}$ can be uniquely expressed as $\textbf{r}=\textbf{r}_1e_1+\textbf{r}_2e_2+\textbf{r}_3e_3+\textbf{r}_4e_4$, where $\textbf{r}_j=(r_{0j},r_{1j},\cdots,r_{n-1,j})\in Z_4^n$ and $j=1,2,3,4$. For any $\textbf{r},~\textbf{s}\in R^n$, we get
 $$\textbf{r}\cdot \textbf{s} = (\textbf{r}_1\cdot \textbf{s}_1)e_1+(\textbf{r}_2\cdot \textbf{s}_2)e_2+(\textbf{r}_3\cdot \textbf{s}_3)e_3+(\textbf{r}_4 \cdot \textbf{s}_4)e_4,$$
 where $\textbf{s}=\textbf{s}_1e_1+\textbf{s}_2e_2+\textbf{s}_3e_3+\textbf{s}_4e_4$, $\textbf{s}_j= (s_{0j},s_{1j},\cdots,s_{n-1,j})\in\mathbb{Z}_4^n$, and $\textbf{r}_j\cdot \textbf{s}_j=\sum_{k=0}^{n-1}{r_{kj}s_{kj}}$.

 Let $C$ be a linear code over $R$, we denote $C_i~(1\le i\le 4)$ as:
 \begin{align*}
C_1&=\{ \textbf{a}\in \mathbb{Z}_4^n|\exists ~\textbf{b},\textbf{c},\textbf{d}\textbf{} \in \mathbb{Z}_4^n,\textbf{a}e_1+ \textbf{b}e_2+\textbf{c}e_3+\textbf{d}e_4\in C\},\\
C_2&=\{ \textbf{b}\in \mathbb{Z}_4^n|\exists ~\textbf{a},\textbf{c},\textbf{d }\in \mathbb{Z}_4^n,\textbf{a}e_1+ \textbf{b}e_2+\textbf{c}e_3+\textbf{d}e_4\in C\},\\
C_3&=\{ \textbf{c}\in \mathbb{Z}_4^n|\exists ~\textbf{a},\textbf{b},\textbf{d} \in \mathbb{Z}_4^n,\textbf{a}e_1+ \textbf{b}e_2+\textbf{c}e_3+\textbf{d}e_4\in C\},\\
C_4&=\{ \textbf{d}\in \mathbb{Z}_4^n|\exists ~\textbf{a},\textbf{b},\textbf{c} \in \mathbb{Z}_4^n,\textbf{a}e_1+ \textbf{b}e_2+\textbf{c}e_3+\textbf{d}e_4\in C\}.
\end{align*}

 Clearly, $C_i~(1\le i\le 4)$ is a linear code of length $n$ over $\mathbb{Z}_4$. And $C$ can be uniquely expressed as $C=e_1C_1+e_2C_2+e_3C_3+e_4C_4$. According to the direct sum decomposition in above, we have $|C|=|C_1| |C_2||C_3||C_4|$. Furthermore, we have\\

\textit{\noindent\textbf{Theorem 1.} Let $C$ be a linear code of length $n$ over $R$, then}

(1) \textit{$C=e_1C_1+e_2C_2+e_3C_3+e_4C_4$, where $C_i~(1\le i\le 4)$ is a linear code of length $n$ over $\mathbb{Z}_4$, and the direct sum decomposition is unique.}

(2) \textit{$C^ \bot=e_1C_1^\bot+e_2C_2^\bot+e_3C_3^\bot+e_4 C_4^\bot$, where $C_i^\bot$ is the dual code of $C_i~(1\le i\le 4)$.}

(3) \textit{$C$ is a self-orthogonal code if and only if $C_i~(1\le i\le 4)$ is a self-orthogonal code over $\mathbb{Z}_4$. Furthermore, $C$ is a self-dual code if and only if $C_i~(1\le i\le 4)$ is a self-dual code over $\mathbb{Z}_4$.\\}

\noindent\textbf{Proof.} (1) It is easily verified by the decomposition in above.\\
(2) Let $D=e_1C_1^\bot+e_2C_2^\bot+e_3C_3^\bot+e_4C_4^\bot$, for any $\textbf{c}\in C,\textbf{d}\in D$, then $\textbf{c}\cdot \textbf{d}=\sum\limits_{i=1}^4{(\textbf{c}_i\cdot \textbf{d}_i )e_i}$, where $\textbf{c}=\textbf{c}_1e_1+\textbf{c}_2 e_2+\textbf{c}_3e_3+\textbf{c}_4e_4$, $\textbf{d}=\textbf{d}_1e_1+\textbf{d}_2e_2+\textbf{d}_3e_3+\textbf{d}_4e_4$, $\textbf{c}_i\in C_i,\textbf{d}_i\in C_i^\bot$. Clearly, $\textbf{c}\cdot \textbf{d}=0$, then $D\subseteq C^\bot$. Furthermore, $$|D|=|{C_1^\bot}||{C_2^ \bot}||{C_3^ \bot}||{C_4^ \bot}|=\frac{4^n}{|{C_1}|}\frac{4^n}{|C_2|}\frac{4^n}{|{C_3}|}\frac{4^n }{|{C_4}|}=\frac{|R|^n}{|C|}=|{C^\bot}|,$$
then we have $C^\bot=D$.\\
(3) $C$ is a self-orthogonal code if and only if $C\subseteq C^\bot$. According to (1) and (2), we have $C\subseteq C^\bot$ if and only if $C_i\subseteq C_i^\bot~(1\le i\le 4)$, then $C_i~(1\le i\le 4)$ is a self-orthogonal code over $\mathbb{{Z}}_4$.
Similarly, $C$ is a self-dual code if and only if $C_i~(1\le i\le 4)$ is a self-dual code over $\mathbb{Z}_4$.\qed\\

Following from Theorem 1, we have \\

\textit{\noindent\textbf{Corollary 2.} There are self-dual codes of arbitrary lengths over $R$.\\}

\noindent\textbf{Proof.} It follows from Theorem 1 that there exists a self-dual code over $R$ if and only if there exists a self-dual code over $\mathbb{Z}_4$. Obviously, there exists a self-dual code $\mathbb{Z}_4$ generated by $$
\left({\begin{array}{*{20}c}
   2 & {} & {}  \\
   {} &  \ddots  & {}  \\
   {} & {} & 2  \\
\end{array}}\right)_{n\times n}.$$\qed\\

Furthermore, we give the generator matrix of the linear codes over $R$.

Let $C=e_1C_1+e_2C_2+e_3C_3+e_4C_4$, for $C_i~(1\le i\le 4)$ is a linear code over $\mathbb{Z}_4$, then $C_i$ is permutation-equivalent to a code generated by
\[G_i= \begin{pmatrix}
   I_{k_{i1}} & A_i & B_i  \\
   0 & 2I_{k_{i2}} & {2C_i}  \\
\end{pmatrix}\]
Thus, $C$ is permutation-equivalent to a linear code generated by
\[G= \begin{pmatrix}
 e_1 G_1  \\
 e_2 G_2  \\
 e_3 G_3  \\
 e_4 G_4  \\
 \end{pmatrix}\]
The dual code $C_i^{\perp}$ of the $Z_{4}$-linear code $C_i$ has generator matrix
\[G_i^{\prime}= \begin{pmatrix}
   -B_{i}^{t}-C_{i}^{t}A_{i}^{t} & C_{i}^{t} & I_{n-k_{i1}-k_{i2}}  \\
   2A_{i}^{t} & 2I_{k_{i2}} & 0  \\
\end{pmatrix}\]
Then $C^{\perp}$ is permutation-equivalent to a linear code generated by
\[H= \begin{pmatrix}
 e_1 G_1^{\prime}  \\
 e_2 G_2^{\prime}  \\
 e_3 G_3^{\prime}  \\
 e_4 G_4^{\prime}  \\
 \end{pmatrix}\]
$H$ is called the check matrix of $C$.\\

Now we study some properties of the linear codes over $R$, which is about the Gray images as following.
From the definition of the Gray map and the Lee weight over $R$, we have $\Phi$ is a distance preserving map from $R^{n}$ to $\mathbb{Z}^{4n}$. Let $C$ be a linear code of length $n$ over $R$, If $\textbf{c}=\textbf{c}_1e_1+\textbf{c}_2e_2+\textbf{c}_3e_3+\textbf{c}_4e_4\in C$, Then $\Phi(\textbf{c})=(\textbf{c}_1,\textbf{c}_2,\textbf{c}_3,\textbf{c}_4)\in Z_4^{4n}$.
 
If $A,B,C,D$ are four codes of length $n$ over $\mathbb{Z}_4$, we define $A\otimes B\otimes C\otimes D =\{(\textbf{a},\textbf{b},\textbf{c},\textbf{d}):\textbf{a} \in A,~\textbf{b}\in B ,~\textbf{c} \in C,~\textbf{d} \in D\}$. Thus we have,\\

\textit{\noindent\textbf{Theorem 3.} Let $C=e_1C_1+e_2C_2+e_3C_3+e_4C_4$ be a linear code of length $n$ over $R$, then} \begin{center}
\textit{$\Phi(C)=C_1\otimes C_2\otimes C_3\otimes C_4$ and $\Phi(C)^\bot=\Phi(C^\bot)$.}
\end{center}
\textit{If $C$ is a self-dual code, then $\Phi(C)$ is also a self-dual code.\\}

\noindent\textbf{Proof.} Firstly, $C_1\otimes C_2\otimes C_3\otimes C_4 \subseteq \Phi(C)$. Here $|C_1\otimes C_2\otimes C_3\otimes C_4|=|C_1||C_2||C_3||C_4|=|C|$, which means $\Phi(C)=C_1\otimes C_2\otimes C_3\otimes C_4$. 

From Theorem 1 (2) we have $\Phi(C^\bot)=C_1^\bot\otimes C_2^\bot \otimes C_3^\bot \otimes C_4^\bot,$ hence $|\Phi(C^\perp)|=\frac {4n}{|C|}.$

Let $\textbf{c}=\textbf{c}_1e_1+\textbf{c}_2e_2+\textbf{c}_3e_3+\textbf{c}_4e_4\in C$, $\textbf{d}=\textbf{d}_1e_1+\textbf{d}_2e_2+\textbf{d}_3e_3+\textbf{d}_4e_4\in C^\bot$, then $\Phi(\textbf{c})\cdot \Phi(\textbf{d})=\sum_{i=1}^4(\textbf{c}_i\cdot \textbf{d}_i)$, which means then then $\Phi(C)^\bot\supseteq\Phi(C^\bot )$. Furthermore, $|{\Phi(C)^\bot}|=\frac{4n}{\Phi(C)}=|\Phi(C^\perp)|$, hence $\Phi(C)^\bot=\Phi(C^\bot)$. \qed\\

Let $G_{i}~(0\leq i\leq 4)$ be the generator matrix of $C_{i}$. From Theorem 3 we have the generator matrix of $\Phi(C)$ is
\[ \begin{pmatrix}
\begin{matrix}G_1 \end{matrix}
&\text{\Large 0}&\text{\Large 0}&\text{\Large 0}\\
\text{\Large 0}&\begin{matrix}G_2 \end{matrix}
&\text{\Large 0}&\text{\Large 0}\\
\text{\Large 0}&\text{\Large 0}& \begin{matrix}G_3 \end{matrix}
&\text{\Large 0}\\
\text{\Large 0}&\text{\Large 0}&\text{\Large 0}& \begin{matrix}G_4 \end{matrix}
\end{pmatrix}.
\]\\

\dse{4~~The MacWilliams identities}

First, we classify elements of $R$ into $D_{0},D_{1},D_{2},D_{3},D_{4},D_{5},D_{6},D_{7},D_{8}$ , where
$D_{0}=\{0\}$ \\
$D_{1}=\{uv,3u+uv,3uv,u+uv,v+3uv,3+u+v+3uv,3v+uv,1+3u+3v+uv\}$ \\
$D_{2}=\{1+3u,3+u,u,3u,2uv,u+2uv,3u+2uv,2u+2uv,v,3+u+v,3u+v,3+v,u+2uv,3u+v+2uv,u+v+2uv,3+2u+v+2uv,2v+2uv,1+3u+2v+2uv,3+u+2v+2uv,
2+2u+2v+2uv,3v,1+3u+3v,u+3v,1+3v,3v+2uv,1+3u+3v+2uv,3u+3v+2uv,1+2u+3v+2uv\}$\\
$D_{3}=\{1+3u+uv,3+u+uv,u+uv,2u+uv,1+2u+uv,3+uv,\cdots,u+3u+3uv,3u+3v+3uv,2u+3v+3uv,1+3u+3uv,1+2u+3v+3uv,1+u+3v+3uv\}$\\
$D_{5}=\{2+2u+uv,1+uv,3+2u+uv,3+3u+uv,1+2u+uv,3+2u+uv,\cdots,2+2u+3v+3uv,3+2u+3v+3uv,3+3u+3v+3uv,2+u+3v+3uv,2+3v+3uv\}$\\
$D_{6}=\{1+u,3+3u,2+u,2+3u,2+2u+2uv,2+u+2uv,2+2uv,1+v,1+u+v,2+3u+v,2+v,1+2u+v+2uv,1+u+v+2uv,2+u+v+2uv,2+v+2uv,2u+2v++2uv,1+u+2v+2uv,
3+3u+2v+2uv,2+2v+2uv,3+3v,3+3u+3v,2+u+3v,2+3v,3+2u+3v+2uv,3+3u+3v+2uv,2+3u+3v+2uv,2+3v+2uv\}$\\
$D_{7}=\{2+uv,,2+3u+uv,2+u+3uv,2+3uv,1+u+v+3uv,2+v+3uv,3+3u+3v+uv,2+3v+uv\}$\\
$D_{8}=\{2\}$\\
$D_4=R/{(D_0\cup D_1\cup D_2\cup D_3\cup D_5\cup D_6\cup D_7\cup D_8 )}$.\\
The elements in $D_{i}$ with same Lee weight, and if $r\in D_{i}$, we have $w_L(r)=i~(0\le i\le 8)$. Let the elements of $R$ be represented by $a_{1},a_{2},\cdots,a_{256}$ according to the sequence of elements in $D_{i}$.

Define $D_iD_j=\{{xy|{x\in D_i,y\in D_j}}\}$, we have $D_0D_j=D_0~(0\leq j\leq 8)$ and
$|D_0|=|D_8|=1,|D_1|=|D_7|=8,|D_2|=|D_6|=28,|D_3|=|D_5|=56,|D_4|=70$.

Let $I$ be a non-zero ideal of $R$. Define $\chi:I\rightarrow\mathbb{C}^{\ast}$ by $\chi(a+ub+vc+uvd)=i^{d}$, where $\mathbb{C}^{\ast}$ is the multiplicative group of unit complex numbers. $\chi$ is a non-trivial character of $I$.

Let $C$ be a linear code of length $n$ over $R$. Define the Hadamard Tranform $\hat f(\textbf{c})=\sum_{\textbf{d}\in R^n}{\chi(\textbf{c}\cdot \textbf{d})f(\textbf{d})}$. We have
$\sum_{\textbf{d}\in C^\bot}{f(\textbf{d})}=\frac{1}{|C|}\sum_{\textbf{c}\in C}{\hat f(\textbf{c})}$, which is noted by $(\ast)$.

The complete Lee weight enumerator (clwe) of a linear code $C$ over $R$ is defined as $$
clwe_C(x_1,x_2,\cdots,x_{256})=\sum_{\textbf{c}\in C}{x_1^{wt_{a_1}(\textbf{c})}}x_2^{wt_{a_2}(\textbf{c})}\cdots x_{256}^{wt_{a_{256}}(\textbf{c})},$$
where $wt_{a_i}(\textbf{c})$ is the number of $a_{i}$ in $\textbf{c}$. This is a homogeneous polynomial in 256 variables $x_{1},x_{2},\cdots,x_{256}$ with total degree on each term being $n$, the length of $C$.\\

\textit{\noindent\textbf{Theorem 4.} Let $C$ be a linear code of length $n$ over $R$. Then
\[clwe_{C^\bot}(x_1,x_2,\cdots,x_{256})=\frac{1}{{\left|C\right|}}clwe_C(M(x_1,x_2,\cdots,x_{256})^T),\] where $M$ is an $|R|\times|R|$ matrix defined by $M(i,j)=\chi(a_ia_j)$.\\}

\noindent\textbf{Proof.} Let $f(x)=\prod_{i=1}^{256}{x_i^{wt_{a_i}(x)}}$. The result follows from the definition of the complete Lee weight enumerator and $(\ast)$ in above.\qed\\

Permutation equivalent codes have the same complete Lee weight enumerators but equivalent codes may have distinct weight enumerators. So the appropriate weight enumerator for studying equivalent codes is the symmetrized Lee weight enumerator(slwe), defined as
\begin{align}
&slwe_C(x,y,z,w,p,s,t,l,m)=clwe_C(x,\underbrace {y,\cdots,y}_8,\underbrace{z,\cdots,z}_{28},\underbrace{w,\cdots ,w}_{56},\underbrace{p,\cdots,p}_{70},\underbrace {s, \cdots ,s}_{56},\underbrace {t, \cdots ,t}_{28},\nonumber\\
&\underbrace{l,\cdots,l}_8,m).\nonumber
\end{align}
Where $x,y,x,w,p,s,t,m,l$ represent the elements of weight 0,1,2,3,4,5,6,7,8, respectively. Therefore, $$
slwes(x,y,z,w,p,s,t,l,m)=\sum_{\textbf{c}\in C}{x^{wt_0(\textbf{c})}y^{wt_1(\textbf{c})}z^{wt_2(\textbf{c})}w^{wt_3(\textbf{c})}p^{wt_4(\textbf{c})}}s^{wt_5(\textbf{c})}t^{wt_6 (\textbf{c})}l^{wt_7(\textbf{c})}m^{wt_8(\textbf{c})},$$
where \\$wt_{0}=wt_{a_{1}}(\textbf{c})$,$wt_1(\textbf{c})=\sum_{i = 2}^9{wt_{a_i }(\textbf{c})}$,$wt_2(\textbf{c})=\sum_{i=10}^{37}{wt_{a_i }(\textbf{c})}$,$wt_3(\textbf{c})=\sum_{i=38}^{93} {wt_{a_i}(\textbf{c})}$,$wt_4(\textbf{c})=\sum_{i=94}^{163}{wt_{a_i }(\textbf{c})}$,
 $wt_5(\textbf{c})=\sum_{i=164}^{219}{wt_{a_i}(\textbf{c})}$,
$wt_6(\textbf{c})=\sum_{i=220}^{247}{wt_{a_i}(\textbf{c})}$,
$wt_7(\textbf{c})=\sum_{i=248}^{255}{wt_{a_i}(\textbf{c})}$,
$wt_8(\textbf{c})=wt_{a_{256}}(\textbf{c})$.\\

\textit{\noindent\textbf{Theorem 5.} Let $C$ be a linear code of length $n$ over $R$. Then $$
slwe_{C^ \bot}(x,y,z,w,p,s,t,l,m)=\frac{1}{|C|}slwe_C(B_0,B_1,B_2,B_3,B_4,B_5,B_6,B_7,B_8),$$ where
\begin{align}
B_0 &=x+8y+28z+56w+70p+56s+28t+8l+m,\nonumber\\
B_1 &=x+6y+14z+14w-14s-14t-6l-m,\nonumber\\
B_2 &=x+4y+4z-4w-10p-4s+4t+4l+m,\nonumber\\
B_3 &=x+2y-2z-6w+6s+2t-2l-m,\nonumber\\
B_4 &=x-4z+6p-4t+m,\nonumber\\
B_5 &=x-2y-2z+6w-6s+2t+2l-m,\nonumber\\
B_6 &=x-4y+4z+4w-10p+4s+4t-4l+m,\nonumber\\
B_7 &=x-6y+14z-14w+14s-14t+6l-m,\nonumber\\
B_8 &=x-8y+28z-56w+70p-56s+28t-8l+m.\nonumber
\end{align}}\\
\noindent\textbf{Proof.}  If $i\in D_0$, then $\sum_{r\in D_0}{\chi(ir)=1}$,
$\sum_{r\in D_1}{\chi(ir)=8}$,$\sum_{r\in D_2}{\chi(ir)=28}$,
$\sum_{r\in D_3}{\chi(ir)=56}$,\\$\sum_{r\in D_4}{\chi(ir)=70}$,
$\sum_{r\in D_5}{\chi(ir)=56}$,$\sum_{r\in D_6}{\chi(ir)=28}$,
$\sum_{r\in D_7}{\chi(ir)=8}$,$\sum_{r\in D_8}{\chi(ir)=1}$. If $i\in D_1$, then
$\sum_{r\in D_0}{\chi(ir)=1}$,$\sum_{r\in D_1}{\chi(ir =6}$,
$\sum_{r\in D_2}{\chi(ir)=14}$,$\sum_{r\in D_3}{\chi(ir)=14}$,
$\sum_{r\in D_4}{\chi(ir)=0}$,\\$\sum_{r\in D_5}{\chi(ir)=-14}$,
$\sum_{r\in D_6}{\chi(ir)=-14}$,$\sum_{r\in D_7}{\chi(ir)=-6}$,
$\sum_{r\in D_8}{\chi(ir)=-1}$. Others can be got similarly.\\
Furthermore,
\begin{align}
&slwe_{C^\bot}(x,y,z,w,p,s,t,l,m)\nonumber\\ &=
clwe_{C^\bot}(x,\underbrace{y,\cdots,y}_{8},\underbrace{z,\cdots,z}_{28},\underbrace{w,\cdots,w}_{56},\underbrace{p,\cdots,p}_{70},
\underbrace{s,\cdots,s}_{56},\underbrace{t,\cdots,t}_{28},\underbrace{l,\cdots,l}_{8}, m)\nonumber\\
&=\frac{1}{|C|}clwe_C(\sum_{i=0}^8{\sum_{r\in D_i }{\chi(a_1 r)}}X_i,\sum_{i=0}^8{\sum_{r\in D_i}{\chi(a_2 r)}}X_i,
\cdots,\sum_{i=0}^8{\sum_{r\in D_i}{\chi(a_{256}r)}}X_i)\nonumber,
\end{align}
where $X_i(0\le i\le 8)$ represents $x,y,z,w,p,s,t,l,m$ respectively. If both $a_j$ and $a_k$ are in $D_i$, we have $$
\sum_{i=0}^8{\sum_{r\in D_i} {\chi (a_j r)}}X_i=\sum_{i=0}^8{\sum_{r\in D_i }{\chi (a_k r)}} X_i,$$ and
\begin{align*}
&slwe_{C^\bot}(x,y,z,w,p,s,t,l,m)\nonumber\\
&=\frac{1}{|C|}slwe_C(\sum_{i=0}^8{\sum_{r \in D_i }{\chi(a_1 r)}}X_i,\sum_{i=0}^8{\sum_{r\in D_i}{\chi(a_2 r)}}X_i , \cdots,\sum_{i=0}^8{\sum_{r\in D_i}{\chi(a_{256}r)}}X_i).
\end{align*}

After calculating, we have $$
\sum_{i=0}^8{\sum_{r \in D_i }{\chi (a_1 r)}}X_i=B_0,\sum_{i=0}^8{\sum_{r\in D_i}{\chi(a_2 r)}} X_i=B_1,\cdots,\sum_{i=0}^8 {\sum_{r\in D_i}{\chi(a_{256}r)}}X_i=B_8.$$
Thus $slwe_{C^\bot}(x,y,z,w,p,s,t,l,m)=\frac{1}{|C|}slwe_C(B_0,B_1,B_2,B_3,B_4,B_5,B_6,B_7,B_8).$\qed\\

Let $C$ be a linear code of length $n$ over $R$ and $A_{i}$ be the number of elements of the Lee weight $i$ in $C$. Then the set $
{A_0,A_1,\cdots,A_{8n}}$ is called the Lee weight distribution of $C$. Define the Lee weight enumerator of $C$ as
$Lee_C(x,y)=\sum_{i=0}^{8n}{A_i x^{8n-i} y^i}$, and $Lee_C(x,y)=\sum_{c\in C}{x^{8n-wt_L(c)}y^{wt_L(c)}}$.\\

\textit{\noindent\textbf{Theorem 6.} Let $C$ be a linear code of length $n$ over $R$. Then $$
Lee_C(x,y)=slwe_C(x^8,x^7y,x^6y^2,x^5y^3,x^4y^4,x^3y^5,x^2y^6,x^1y^7,y^8).$$\\}

\noindent\textbf{Proof.} Let $wt_L(c)=\sum_{i=0}^8{iwt_i(\textbf{c})}$. For $n=\sum_{i=1}^{256}{wt_{a_i}}(\textbf{c})=\sum_{i=0}^8{wt_i(\textbf{c})}$, we have
$$8n-wt_L(\textbf{c})=\sum_{i=0}^8{(8-i)wt_i(\textbf{c})}.$$ From the definition of the Lee weight enumerator of $C$ in above,
we have $$
Lee_C(x,y)=\sum_{\textbf{c}\in C}{x^{8n-wt_L(\textbf{c})}y^{wt_L(\textbf{c})}}=\sum_{\textbf{c}\in C}{x^{\sum_{i=0}^8{(8-i)wt_i(\textbf{c})}} y^{\sum_{i=0}^8{iwt_i(\textbf{c})}}}
=\sum_{\textbf{c}\in C}{\prod_{i=0}^8{(x^{8-i}y^i)^{wt_i(\textbf{c})}}}$$
$=slwe_C(x^8,x^7y,x^6y^2,x^5y^3,x^4y^4,x^3y^5,x^2y^6,x^1y^7,y^8).$\qed\\

\textit{\noindent\textbf{Theorem 7.} Let $C$ be a linear code of length $n$ over $R$. Then $$
Lee_{C^\bot}(x,y)=\frac{1}{|C|}Lee_C(x+y,x-y).$$\\}

\noindent\textbf{Proof.} From Theorem 5 and Theorem 6, we have $$
Lee_{C^\bot}(x,y)=\frac{1}{|C|}slwe_C(E_0,E_1,E_2,E_3,E_4,E_5,E_6,E_7,E_8),$$ where\\
$E_0=x^8+8x^7y+28x^6y^2+56x^5y^3+70x^4y^4+56x^3y^5+28x^2y^6+8xy^7+y^8=(x+y)^8$,\\
$E_1=x^8+6x^7y+14x^6y^2+14x^5y^3-14x^3y^5-14x^2y^6-6xy^7-y^8=(x+y)^7(x-y)$,\\
$E_2=x^8+4x^7y+4x^6y^2-4x^5y^3-10x^4y^4-4x^3y^5+4x^2y^6+4xy^7+y^8=(x+y)^6(x-y)^2$,\\
$E_3=x^8+2x^7y-2x^6y^2-6x^5y^3+6x^3y^5+2x^2y^6-2xy^7-y^8=(x+y)^5(x-y)^3$,\\
$E_4=x^8-4x^6y^2+6x^4y^4-4x^2y^2+y^8=(x+y)^4(x-y)^4$,\\
$E_5=x^8-2x^7y-2x^6y^2+6x^5y^3-6x^3y^5+2x^2y^6+2xy^7-y^8=(x+y)^3(x-y)^5$,\\
$E_6=x^8-4x^7y+4x^6y^2+4x^5y^3-10x^4y^4+4x^3y^5+4x^2y^6-4xy^7+y^8=(x+y)^2(x-y)^6$,\\
$E_7=x^8-6x^7y+14x^6y^2-14x^5y^3+14x^3y^5-14x^2y^6+6xy^6-y^8=(x+y)(x-y)^7$,\\
$E_8=x^8-8x^7y+28x^6y^2-56x^5y^3+70x^4y^4-56x^3y^5+28x^2y^6-8xy^7+y^8=(x-y)^8$.\\
Thus $Lee_{C^\bot}(x,y)=\frac{1}{|C|}Lee_C(x+y,x-y)$.\qed\\

\dse{5~~Cyclic codes over $R$}
Cyclic codes are an important class of linear codes. They have been studied over many rings. In this section, we discuss cyclic codes over the ring $R$.\\

\textit{\noindent\textbf{Theorem 8.} Let $C=e_1C_1+e_2C_2+e_3C_3+e_4C_4$. Then $C$ is a cyclic code over $R$ if and only if one of following three conditions is satisfied:\\
(1) $C_i~(1\leq i\leq 4)$ is a cyclic code over $\mathbb{Z}_{4}$.\\
(2) $C_i^\bot~(1\leq i\leq 4)$ is a cyclic code over $\mathbb{Z}_{4}$.\\
(3) $C^\bot$ is a cyclic code over $R$.\\}

\noindent\textbf{Proof.} For any $c_i=(c_{i0},c_{i1},\cdots,c_{i,n-1})\in C_i~(1\leq i\leq 4)$, then $c=e_1c_1+e_2c_2+e_3c_3+e_4c_4 \in C$. Since $C$ is a cyclic code, we have
\[d=(\sum_{i=1}^4{e_ic_{i,n-1}},\sum_{i=1}^4{e_ic_{i0}},\cdots,\sum_{i = 1}^4{e_ic_{i,n-2}})\in C,\]
then $(c_{i,n-1},c_{i0},\cdots,c_{i,n-2})\in C_i$. Thus $C_{i}$ is a cyclic code over $\mathbb{Z}_{4}$. Vice versa.

Since $C_{i}$ is a cyclic code over $\mathbb{Z}_{4}$, we have $C_i^\bot$ is a cyclic code over $\mathbb{Z}_{4}$. From (1) we have $C^\bot$ is a cyclic code over $R$. Further, $C^\bot$ is a cyclic code over $R$.\qed\\

Let $R_n=R[x]/(x^n-1)$. Obviously, cyclic codes of length $n$ over $R$ are precisely ideals of $R_{n}$. We will use the generator polynomial of $C_{i}$ over $\mathbb{Z}_{4}$ to construct the generator polynomials of cyclic codes over $R$. Cyclic codes over $\mathbb{Z}_{4}$ have following results.\\

\textit{\noindent\textbf{Lemma 9[2,3].} Let $n$ be an odd positive integer, and $x^n-1=\prod_{i=1}^r{f_i(x)}$ be the unique factorization of $x^{n}-1$, where $f_1 (x),\cdots,f_r(x)$ are basic irreducible polynomials over $Z_{4}$. Let $C$ be a cyclic code of length $n$ over $\mathbb{Z}_{4}$, then
$$C=(f_0(x),2f_1(x))=(f_0(x)+2f_1(x))$$
where $f_0(x)$ and $f_1(x)$ are the monic factors of $x^n-1$ and $f_1(x)|f_0(x)$.\\}

Generally, if a linear code of any length $n$ over $\mathbb{Z}_{4}$, then there exist monic polynomials $f(x),g(x),p(x)\in Z_4[x]$ such that
$$C=(f(x)+2p(x),2g(x))$$
where $g(x)|f(x)|(x^n-1), g(x)|p(x)(\frac{x^n-1}{f(x)})$ and $|C|=2^{2n-\deg(f(x))-\deg(g(x))}.$\\

\textit{\noindent\textbf{Theorem 10.} Let $C=e_1C_1+e_2C_2+e_3C_3+e_4C_4$ be a cyclic code of any length $n$ over $R$, there exist $f_i(x),g_i(x),p_i(x)\in \mathbb{Z}_4[x](1\leq i\leq 4)$, such that $C_i=(f_i(x)+2p_i(x),2g_i(x))$, then $C=(\sum_{i=1}^4{e_if_i}(x)+2 \sum_{i=1}^4{e_ip_i(x)}, 2 \sum_{i=1}^4{e_ig_i (x)})$. Further, if $n$ is odd, then $C= (\sum_{i=1}^4e_i(f_i(x)+2g_i(x)).$ \\}

\noindent\textbf{Proof.} Let $D=( \sum_{i=1}^4{e_if_i }(x)+2 \sum_{i=1}^4{e_ip_i(x)}, 2 \sum_{i=1}^4{e_ig_i (x)})$. For any $c(x)\in C$, there exist $u_i(x),v_i(x)\in \mathbb{Z}_4[x]$ such that $$c(x)=\sum_{i=1}^4{e_i((f_i(x)+2p_i(x))u_i(x)+g_i(x)v_i (x))}.$$
But
\begin{align*}
&\sum_{i=1}^4{e_i((f_i(x)+2p_i(x))u_i(x)+g_i(x)v_i(x))}\\
&=\sum_{i=1}^4{e_i u_i(x)}\sum_{i=1}^4{e_i(f_i(x)}+2p_i(x))+\sum_{i=1}^4{e_1v_1(x)}\sum_{i=1}^4{e_ig_i(x)},
\end{align*}
then $C\subseteq D$. Obviously, $D\subseteq C$. Thus $C=D$.\qed\\

Next, we start to study the generator polynomial of the dual code of $C$. For any $f(x)|(x^n-1)$,
let $\hat{f}(x)=\frac{x^n-1}{f(x)}$. Define the reciprocal polynomial of $f(x)$ by $f(x)^*=x^{deg(f)}f(x^{-1})$ and the annihilator of $C$ by $Ann(C)=\{c'|c\cdot c'=0,c\in C\}$. Let $C$ is a cyclic code of length $n$ over $Z_4$, then $C=(f(x)+2p(x),2g(x))$,
 where $g(x)|f(x)|(x^n-1)$,~$\deg(p(x))<\deg(g(x))$, and $p(x)\frac{{x^n-1}}{{f(x)}}=g(x)u(x)$.\
Thus
$$( \hat{g}(x)+2u(x))(f(x)+2p(x))=2(f(x)u(x)+ \hat{g}(x)p(x))=0$$
and $ deg(u(x))< deg(\hat{g}(x))$.

Thus we have\\

\textit{\noindent\textbf{Theorem 11.} Let $C=(f(x)+2p(x),2g(x))$ be a cyclic code of length $n$ over $\mathbb{Z}_{4}$, then $$C^\bot=(\hat{g}(x)^*+2x^{deg(\hat{g}(x))-deg(u(x))}u(x)^*,2\hat{f}(x)^*).$$\\}

\noindent\textbf{Proof.} Let $D=(\hat{g}(x)+2u(x),2\hat{f}(x))$. Since $(\hat{g}(x)+2u(x))(f(x)+2p(x))=0$, we have $D \subseteq Ann(C)$. And $|D|=2^{2n-deg(\hat{g}(x))-deg(\hat{f}(x))}=2^{deg(g(x))+deg(f(x))}=|Ann(C)|$, thus $D=Ann(C)$. But $C^\bot=Ann(C)^*=(\hat{g}(x)^*+2x^{deg(\hat{g}(x))-deg(u(x))}u(x)^*,2\hat{f}(x)^*)$.

Specially, if $n$ is odd and $g(x)|f(x)|(x^n-1)$, then $(g(x),\hat{f}(x))=1$. We can verify that $p(x)=0=u(x)$. So we have $C=(f(x),2g(x))$, therefore $C^\bot=(\hat{g}(x)^{*},2\hat{f}(x)^{*})$.

Thus we get the generator polynomial of the dual code $C^\bot$ over $R$ and $C^\bot=(\sum_{i=1}^4{e_i\hat{g}_i(x)^*}+2\sum_{i=}^4{e_ix^{deg(\hat{g}_i(x))-deg(u_i(x))}u_i(x)^*},2\sum_{i=1}^4{e_i\hat{f}_i(x)^*})$.\qed\\

Let $\sigma$ be the cyclic shift over $Z_{4}$. For any positive integer $s$, let $\sigma_{s}$ be the quasi-shift given by
 $$\sigma_{s}(a^{(1)}\mid a^{(2)}\mid \cdots\mid a^{(s)})=\sigma(a^{(1)}) \mid \sigma(a^{(2)})\cdots\mid\sigma(a^{(s)}),$$
where $a^{(1)},a^{(2)},\cdots,a^{(s)}\in Z_{4}^{n}$ and $"\mid"$ denotes the usual vector concatenation. A quaternary quasi-cyclic code $C$ of index $s$ and length $ns$ is a subset of $(Z_{4}^{n})^{s}$ such that $\sigma_{s}(C)=C$.\\

\textit{\noindent\textbf{Theorem 12.}  Let $C=C_1e_1+C_2e_2+C_3e_3+C_4e_4$ be a cyclic code of length $n$ over $R$. Then $\Phi(C)$ is a quasi-cyclic code of index $4$ and length $4n$ over $\mathbb{Z}_{4}$.\\}

\noindent\textbf{Proof.} Let $(c_0,c_1,\cdots,c_{n-1})\in C$. Let $c_{j}=r_{1,j}e_1+r_{2,j}e_2+r_{3,j}e_3+r_{4,j}e_4$, where $r_{i,j}\in D_{i}(0\le i\le 4,0\le j\le n-1)$. Since $C$ be a cyclic code, then $\tau(c_0,c_1,\cdots,c_{n-1})=(c_{n-1},c_0,\cdots,c_{n-2})\in C$ and
$\Phi(c_0,c_1,\cdots,c_{n-1})= (r_{1,0},\cdots,r_{1,n-1},r_{2,0},\cdots,r_{2,n-1},r_{3,0},\cdots,r_{3,n-1},r_{4,0},\cdots,r_{4,n-1})$ and $D_{i}(0\leq i\leq 4)$ is a cyclic code. Then we have
\begin{align*}
&\sigma_{4}(r_{1,0},\cdots,r_{1,n-1}\mid r_{2,0},\cdots,r_{2,n-1}\mid r_{3,0},\cdots,r_{3,n-1}\mid r_{4,0},\cdots,r_{4,n-1})\\
&=\sigma(r_{1,0},\cdots,r_{1,n-1})\mid\sigma(r_{2,0},\cdots,r_{2,n-1})\mid\sigma(r_{3,0},\cdots,r_{3,n-1})
\mid\sigma(r_{4,0},\cdots,r_{4,n-1}))\in \Phi(C).
\end{align*}
 Thus $\Phi(C)$ is a quasi-cyclic code of index $4$ and length $4n$ over $\mathbb{Z}_{4}$.\qed\\

\textit{\noindent\textbf{Theorem 13.} Let $C_{i}~(0\leq i\leq 4)$ be a cyclic code of $n$~($n$ is odd) over $\mathbb{Z}_{4}$ and $C_{i}=(f_{0,i}(x),2f_{1,i}(x))$, where $f_{0,i}(x)$ and $f_{1,i}(x)$ are the monic factors of $x^{n}-1$ over $\mathbb{Z}_{4}$ and $f_{1,i}(x)\mid f_{0,i}(x)$. Then the type of $C_{i}~(0\leq i\leq 4)$ is $4^{n-deg(f_{0,i}(x))}2^{deg(f_{0,i}(x))-deg(f_{1,i})(x)}$.\\}

\noindent\textbf{Proof.} For $C_{i}=(f_{0,i}(x),2f_{1,i}(x))$, the type of $C_{i}~(0\le i\le 4)$ is $4^{k_{0,i}}2^{k_{1,i}}$.  We define a map $'-'$ as: $\mathbb{Z}_{4}\rightarrow \mathbb{Z}_{2}$. Thus $k_{0,i}=dim(\overline{C_{i}})=dim((\overline{f_{0,i}}(x)))=n-deg(f_{0,i}(x))$. Furthermore $\overline{(C_{i}:2)}=\overline{(f_{1,i}(x))}$, thus $k_{0i}+k_{1i}=n-deg(f_{1,i}(x))$. So $k_{1,i}=deg(f_{0,i}(x))-deg(f_{1,i}(x))$. Then the type of $C_{i}~(0\le j\le 4)$ is $4^{n-deg(f_{0,i}(x))}{2^{deg(f_{0,i}(x))-deg(f_{1,i})(x)}}$.\qed\\

Let $C_{i}~(0\leq i\leq 4)$ be a cyclic code of length $n$ over $\mathbb{Z}_{4}$ ,we have \\ 

\textit{\noindent\textbf{Corollary 14.} Let $\Phi(C)=C_1\otimes C_2 \otimes C_3\otimes C_4$ be a linear code of length $4n$~($n$ is odd) over $\mathbb{Z}_{4}$. $C_{i}~(0\le i\le 4)$ be a cyclic code of $n$ over $\mathbb{Z}_{4}$, Then
the type of $\Phi(C)$ is  }
\textit{\begin{center}
$4^{\sum_{i=1}^{4}(n-deg(f_{0,i}(x)))}2^{\sum_{i=1}^{4}(deg(f_{0,i}(x))-deg(f_{1,i})(x))}.$
\end{center}} 

Let $\Phi(C)=C_1\otimes C_2 \otimes C_3\otimes C_4$ be a linear code of length $4n$~($n$ is odd) over $\mathbb{Z}_{4}$ and $d$ be the Lee Distance of $\Phi(C)$. Then $d=min\left\{d_{1},d_{2},d_{3},d_{4}\right\}$, where $d_{i}~(1\le i\le 4)$ is the Lee Distance of $C_{i}$.

Next, we use some cyclic codes of odd length over $R$ to obtain some optimal linear codes of larger length over $\mathbb{Z}_{4}$ by the Gray map. We now consider some cyclic codes over $R$ of different odd lengths and their Gray images and obtain the Table 1.\\

\begin{table}
\caption{ Some cyclic codes of odd length over $R$ and their $\mathbb{Z}_{4}$-images }
\begin{tabular}{ccc}
\toprule
 length $n$ & generator polynomials of $C$  & parameters of $\Phi(C)$\\
\midrule
3 & $x^{2}+x+3$  & $(12,4^{4}2^{8},2)$\\
3 & $(1-u)(x^{2}+x+1)+2$  & $(12,4^{2}2^{10},2)$\\
5 & $x^{4}+x^{3}+x^{2}+x+1$  & $(20,4^{4}2^{0},5)$\\
5 & $(1-v)(x+3)+v(x^{4}+x^{3}+x^{2}+x+1)+2$  & $(20,4^{10}2^{10},2)$\\
7 & $x^{3}+2x^{2}+x+1$  & $(28,4^{16}2^{12},2)$\\
7 & $x^{4}+x^{3}+3x^{2}+3$  & $(28,4^{12}2^{12},4)$\\
7 & $x^{6}+x^{5}+x^{4}+3x^{3}+3x^{2}+x+3$  & $(28,4^{4}2^{12},6)$\\
7 & $x^{4}+x^{3}+3x^{2}+2x+1$  & $(28,4^{0}2^{12},8)$\\
7 & $uvx^{3}+2uvx^{2}+x+1$ & $(28,4^{12}2^{6},2)$\\
7 & $x^{4}+(1+2v)x^{3}+3x^{2}+2x+1$ & $(28,4^{12}2^{8},4)$\\
7 & $(1-uv)(x^{6}+x^{5}+x^{4}+x^{3}+3x^{2}+x+3+uv(x+3)$ & $(28,4^{3}2^{15},4)$\\
7 & $(1-uv)(x^{6}+x^{5}+x^{4}+x^{3}+3x^{2}+x+3+uv(x^{3}+2x^{2}+x+3)$ & $(28,4^{4}2^{12},6)$\\
9 & $(1-u-v+uv)(x^{2}+x+1)+(u+v-uv)(x^{6}+x^{3}+1)+2$ & $(36,4^{16}2^{20},2)$\\
9 & $(1-v)(x+1)+v(x^{8}+x^{7}+x^{6}+x^{5}+x^{4}+x^{3}+x^{2}+x+1)+2$ & $(36,4^{18}2^{18},2)$\\
9 & $ x^{8}+x^{7}+x^{6}+x^{5}+x^{4}+x^{3}+x^{2}+x+3+2uv(x^{6}+x^{3}+3x^{2}+3x)$ & $(36,4^{4}2^{20},4)$\\
\bottomrule
\end{tabular}
\end{table}

\noindent\textbf{Example 1.} Consider a cyclic code over $R$ of length 23. In $\mathbb{Z}_{4}[x]$,
$x^{23}-1=L_{1}(x)L_{2}(x)L_{3}(x)$, where
\begin{align*}
L_{1}(x) &=x+3,\\
L_{2}(x) &=x^{11}+2x^{10}+3x^{9}+3^{7}+3x^{6}+3x^{5}+2x^{4}+x+3,\\
L_{3}(x) &=x^{11}+3x^{10}+2x^{7}+x^{6}+x^{5}+x^{4}+x^{2}+2x+3.
\end{align*}
By Theorem 6, Let $C_{1}=C_{2}=(L_{1}(x)L_{2}(x))+2L_{2}(x))$ and $C_{3}=C_{4}=(L_{1}(x)L_{3}(x)+2L_{3}(x))$ over $\mathbb{Z}_{4}$. By Theorem 6, we have $C$ is a cyclic code and  $C=((1-v)(L_{1}(x)L_{2}(x)+2L_{2}(x))+v(L_{1}(x)L_{3}(x)+2L_{3}(x)))$ over $R$. By Theorem 9, $\Phi(C)$ is a
$(92,4^{44}2^{4},10)$ quaternary quasi-cyclic code over $\mathbb{Z}_{4}$.

\dse{6~~MDS codes over $R$}
We come now to one of the most interesting section in all of coding theory: MDS codes. Let $C$ be a linear code of length $n$ over $R$.

Let $C$ be a linear code of length $n$ over $R$ and $d_{H}$ be the minimum Hamming distance. We have
$|C|\leq |C|^{n-d_{H}+1}$, thus $d_{H}\leq n-{\log_{|R|}}^{|C|}+1$, which is called the Singleton bound. If $C$ meet the Singleton bound, then $C$ is called MDS code.

In paper[20], we have there are only free trival MDS codes over $\mathbb{Z}_{4}$.\\

\textit{\noindent\textbf{Lemma 16.} Let $C$ be a linear code of length $n$ over $\mathbb{Z}_{4}$, then $C$ is a MDS code
if and only if $C$ is either $Z_{4}^{n}$ with parameters$(n,4^{n},1)$,$<1>$ with parameters $(n,4,n)$ or $<1>^{\perp}$ with parameters ~$(n,4^{n-1},2)$, where 1 denote the all-1 vectors.\\}

Further, we discuss linear codes over $R$. Let $C$ be a linear code of length $n$ over $R$ and $C=e_1C_1+e_2C_2+e_3C_3+e_4C_4$, where $C_i(1\le i\le 4)$ is a linear code of length $n$ over $\mathbb{Z}_4$.
$d_{H}$ be the Hamming Distance of $C$. Then $d_{H}=min\left\{d_{H1},d_{H2},d_{H3},d_{H4}\right\}$, where $d_{Hi}~(1\le i\le 4)$ is the Hamming Distance of $C_{i}$. Thus the Singleton bound can be write as
\[ d_{H}\leq n-\frac{1}{{4}}\sum_{1=0}^{4}\log_{4}^{|C_{i}|}+1.\\\]

\textit{\noindent\textbf{Lemma 17.} Let $C$ is a MDS code over $R$.\\
(1)If $d_{H}=1$, then all of $C_{i}$ are MDS codes with parameters $(n,4^{n},1)$ .\\
(2)If $d_{H}=2$, then all of $C_{i}$ are MDS codes with parameters $(n,4^{n-1},2)$.\\}

\noindent\textbf{Proof.} (1)If $d_{H}=1$ and $C$ is a MDS code, then $\sum_{1=0}^{4}\log_{4}^{|C_{i}|}=4n.$
 But $|C_{i}|\leq 4^{n}$, then the identity is true if and only if $|C_{i}|=4^{n}$. Thus $C$ is a $(n,4^{4n},1)$ MDS code if and only if all of $C_{i}$ are $(n,4^{n},1)$ MDS codes.\\
(2)If $d_{H}=2$, then $\sum_{1=0}^{4}\log_{4}^{|C_{i}|}=4(n-1).$ For $d_{H}=min\left\{d_{H1},d_{H2},d_{H3},d_{H4}\right\}$, then $d_{Hi}\geq 2$. By the Singleton bound over $Z_{4}$, we have $|C_{i}|\leq 4^{n-d_{Hi}+1}\leq 4^{n-1}$. Further, we have all of $C_{i}$ are $(n,4^{n-1},2)$ MDS codes.\qed\\

\textit{\noindent\textbf{Theorem 18.} If $C$ is a MDS code over $R$. Then there is at least one $C_{i}~(1\le i\le 4)$ be MDS code.\\}

\noindent\textbf{Proof.} If none of $C_{i}$ is MDS code, then $d_{Hi}< n-\log_{4}^{|C_{i}|}+1.$ For $d_{H}=min\left\{d_{H1},d_{H2},d_{H3},d_{H4}\right\}$, thus $d_{H}< n-\log_{4}^{|C_{i}|}+1$. Further
$d_{H}< n-\frac{1}{{4}}\sum_{1=0}^{4}\log_{4}^{|C_{i}|}+1,$ which is a contradiction.\qed\\

\textit{\noindent\textbf{Theorem 19.} If $C$ is a MDS code over $R$ and there exist three MDS codes of $C_{i}~(1\le i\le 4)$, then the other $C_{i}$  must be MDS code and all of $C_{i}~(1\le i\le 4)$ with same parameters.\\}

\noindent\textbf{Proof.} Without loss of generality, we set $C_{1},C_{2},C_{3}$ be MDS codes, then $d_{Hi}= n-\log_{4}^{|C_{i}|}+1~(1\le i\le 3)$, and $d_{H}=n-\frac{1}{{4}}\sum_{1=0}^{4}\log_{4}^{|C_{i}|}+1.$ Then we have \[4d_{H}-\sum_{i=1}^{3}d_{Hi}=n-\log_{4}^{|C_{4}|}+1\geq d_{H4}\]
Thus $4d_{H}\geq \sum_{i=1}^{4}d_{Hi}$. For $d_{H}=min\left\{d_{H1},d_{H2},d_{H3},d_{H4}\right\}$, then $d_{H1}=d_{H2}=d_{H3}=d_{H4}=d_{H}$.\qed\\

\textit{\noindent\textbf{Theorem 20.} If $C$ is a MDS code over $R$ and there exist two MDS codes of $C_{i}~(1\le i\le 4)$, then the others $C_{i}$  must be MDS code and all of $C_{i}~(1\le i\le 4)$ with same parameters.\\}

\noindent\textbf{Proof.} If $d_{H}=1$ or 2, by Lemma 17, we have $C_{i}~(1\le i\le 4)$ with same parameters. Now let $C_{1},C_{2}$ be MDS codes with same parameters $(n,4,n)$. For $C$ is a MDS code, then $d_{H}= n-\frac{1}{{4}}\sum_{1=0}^{4}\log_{4}^{|C_{i}|}+1$. Further, $|C_{3}||C_{4}|=4^{4n-4d_{H}+2}$. If either $C_{3}$ or $C_{4}$ is MDS code, by Proposition 19, all of $C_{i}~(1\le i\le 4)$ are MDS codes with same parameters. If neither of $C_{3}$ nor $C_{4}$ is MDS code. Given $d_{H3}= d_{H}$, then we have $|C_{3}|<4^{n-d_{H}+1},|C_{4}|>4^{3n-3d_{H}+1}$. For $C_{4}$ is not a MDS code, then we have
\[ d_{H4}<n-\log_{4}^{|C_{4}|}+1<n-3n+3d_{H}-1+1=3d_{H}-2n.\]
and $d_{H}\leq d_{H4}<3d_{H}-2n$, thus $d_{H}>n$. It is a contradiction.\qed\\

From Theorem 18, we have\\

\textit{\noindent\textbf{Theorem 21.}  If $C$ is a MDS code over $R$ and $C_{1}$ is a MDS code with parameters $(n,4,n)$. Then $C_{2},C_{3},C_{4}$ are also MDS codes with parameters $(n,4,n)$.\\}

\noindent\textbf{Proof.} For $C$ is a MDS code over $R$ and $C_{1}$ is a MDS code, then $d_{H}\geq 3$ and
\[\log_{4}^{|C_{2}|}+\log_{4}^ {|C_{3}|}+\log_{4}^{|C_{4}|}=4n-4d_{H}+3.\]
Let $C_{2},C_{3}$ and $C_{4}$ be not MDS codes.
For $d_{H2}\geq d_{H},d_{H3}\geq d_{H},d_{H4}\geq d_{H}$ and $|C_{i}|<4^{n-d_{Hi}+1}\leq 4^{n-d_{H}+1}$, then
we have $|C_{i}|<4^{n-d_{Hi}+1}\leq 4^{n-d_{H}+1}$, i.e. $|C_{i}|\leq 4^{n-d_{H}+1}.$ We have $3n-3d_{H}+3\geq 4n-4d_{H}+3$, thus $d_{H}\geq n$, i.e. $d_{H}= n$. And
\[\log_{4}^{|C_{2}|}+\log_{4}^ {|C_{3}|}+\log_{4}^{|C_{4}|}=3.\]
But $|C_{i}|\leq 4^{n-d_{H}+1}=4$, the equality is true if and only if $|C_{i}|=4$. Then $C_{i}~(1\le i\le 4)$ with same parameters $(n,4,n)$.\qed\\

From the discussion in above, we have\\

\textit{\noindent\textbf{Theorem 22.}  $C$ is a MDS code over $R$ if and only if all of $C_{i}~(1\le i\le 4)$ are MDS codes over $\mathbb{Z}_{4}$ with same parameters.}

\dse{7~~Conclusion}
In this paper, we study the linear codes over $R=\mathbb{Z}_4[u,v]/{(u^2-u,v^2-v,uv-vu)}$. First, we give the decomposition in a direct sum of the ring $R$. Then we use the direct sum decomposition to study the linear codes over $R$, which can be expressed by four relevant linear codes over $\mathbb{Z}_{4}$. Second, we study the MacWilliams identities of linear codes over $R$. Third, we study cyclic codes and the Gray images over $R$. Last, we discuss MDS codes over $R$. Next, we will study the linear codes over $\mathbb{Z}_{p^m}[u,v]/{(u^2-u,v^2-v,uv-vu)}$.

\dse{Acknowledgments}
The authors would like to thank the referees for their helpful comments and a very meticulous reading of this manuscript.

\end{document}